\documentclass[12pt,preprint]{emulateapj}
\newcommand{\kms}{$\rm {km}~\rm s^{-1}$}
\begin{document}
\title{Gemini and $\it{Hubble~Space~Telescope}$ Evidence for an Intermediate Mass Black Hole in omega Centauri} 
\shorttitle{Black Hole in $\omega$ Centauri}
\author{Eva Noyola} 
\affil{Astronomy Department, University of Texas at Austin, Austin, TX 78712}
\affil{Max-Planck-Institut fuer extraterrestrische Physik, 85748, Garching, Germany} 
\author{Karl Gebhardt} 
\affil{Astronomy Department, University of Texas at Austin, Austin, TX 78712} 
\email{noyola@mpe.mpg.de}
\author{Marcel Bergmann} 
\affil{Gemini Observatory, Tucson, AZ 85726}

\begin{abstract}

The globular cluster $\omega$ Centauri is one of the largest and most
massive members of the galactic system. However, its classification as
a globular cluster has been challenged making it a candidate for being
the stripped core of an accreted dwarf galaxy; this together with the
fact that it has one of the largest velocity dispersions for star
clusters in our galaxy makes it an interesting candidate for harboring
an intermediate mass black hole. We measure the surface brightness
profile from integrated light on an $\it{HST}$/ACS image of the
center, and find a central power-law cusp of logarithmic slope
-0.08. We also analyze Gemini GMOS-IFU kinematic data for a
5x5\arcsec~field centered on the nucleus of the cluster, as well as
for a field 14\arcsec~away.  We detect a clear rise in the velocity
dispersion from 18.6~\kms\ at 14\arcsec~to 23 \kms\ in the center. A
rise in the velocity dispersion could be due to a central black hole,
a central concentration of stellar remnants, or a central orbital
structure that is radially biased. We discuss each of these
possibilities. An isotropic, spherical dynamical model implies a black
hole mass of $4.0^{+0.75}_{-1.0}\times10^4 M_\odot$, and excludes the
no black hole case at greater than 99\% significance. We have also run
flattened, orbit-based models and find similar results. While our
preferred model is the existence of a central black hole, detailed
numerical simulations are required to confidently rule out the other
possibilities.

\end{abstract}


\section{Introduction}

\vspace{10pt}

The globular cluster $\omega$ Centauri (NGC~5139) is regarded to be
the largest and most massive member of the Galactic cluster system
with a tidal radius of 69 parsecs \citep{har96}, an estimated mass of
$5.1\times10^6 M_\odot$, and a measured central velocity dispersion of
$22 \pm 4$ \kms\ \citep {mey95}. The cluster presents a large scale
global rotation, measured with radial velocities, of 8 \kms\ at a
radius of 11pc from the center \citep{mer97} and confirmed with proper
motions \citep{van00}, which makes it one of the most flattened
galactic globular clusters \citep{whi87}. A rotating flattened model
including proper motion and radial velocity datasets by \citet{ven06}
calculate a lower total mass of $2.5\times10^6 M_\odot$ and confirm
the central line of sight velocity dispersion value of 20 \kms. They
measure a dynamical distance of $4.8\pm0.3$ kpc (which we adopt for
this paper). $\omega$ Cen has a peculiar highly bound retrograde orbit
\citep{din01}. It also has a stellar population that makes it stand
out from the rest of the Galactic globular clusters due to its
complexity. It shows a broad metallicity distribution
\citep{bed04,nor96}, as well as a kinematical and spatial separation
between the different subpopulations \citep{pan03, nor97}. 

All the above results have led to the hypothesis that $\omega$ Cen is
not a classical globular cluster, but instead is the nucleus of an
accreted galaxy \citep{fre03,bek03,mez05}. The scenarios of it being
the product a merger of two globular clusters \citep{ick88} and of
self-enrichment \citep{iku00} have also been proposed to explain the
stellar populations.

The high measured velocity dispersion together with the possibility of
being a stripped galaxy make $\omega$ Cen an interesting candidate for
harboring a black hole in its center. The extrapolation of
$M_\bullet-\sigma$ relation for galaxies \citep{geb00a,fer00,tre02}
predicts a $1.3\times10^4 M_{\odot}$ black hole for this cluster. The
sphere of influence of such black hole for a star cluster at the
distance of $\omega$ Cen with a velocity dispersion of 20 \kms\ is
$\sim$ 5\arcsec, making it an excellent target for ground-based
observations.

Two globular clusters have been suggested for harboring an
intermediate mass black hole in their nucleus. One is the galactic
cluster M15 \citep{geb00,ger02,ger03} and the other is G1, a giant
globular cluster around M31 \citep{geb02,geb05}. M15 is assumed to be
in a post-core collapse state, therefore its dynamical state has been
debated between harboring a black hole or containing a large number of
compact remnants in its center \citep{bau03a,bau03b}. Unfortunately,
observational constraints between these two hypothesis remain
inconclusive \citep{bos06}. G1 on the other hand, has a core with
characteristics closer to those of $\omega$ Cen, and observations
support the black hole interpretation for G1. \citet{bau03c} propose
an alternative interpretation for G1 in which they match the
observations with a model of two colliding globular clusters. The G1
black hole models are preferred since the M/L profile is expected to
be flat in its core, so any rise in the velocity dispersion is
unlikely to be due to remnants that concentrated there from mass
segregation. The black hole interpretation for G1 is strongly
supported by radio \citep{ulv07} and x-ray \citep {poo06} detections
centered on the nucleus. $\omega$ Cen and G1 have similar properties
in both their photometric and kinematic profiles. In this paper we
report photometric and kinematical measurements that suggest the
presence of a central black hole in $\omega$ Cen.

\begin{figure}
  \plotone{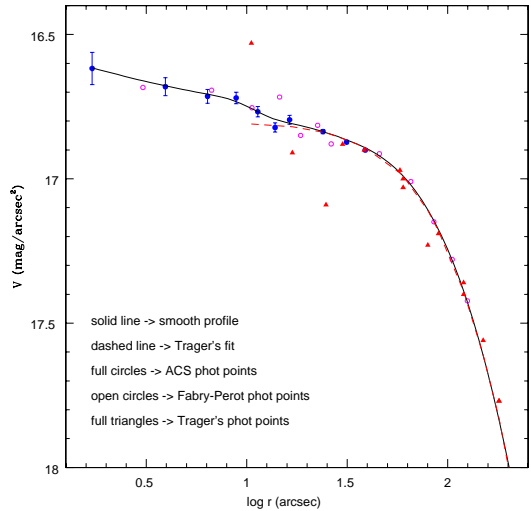}
  \caption{Surface brightness profiles for $\omega$ Cen. The circles
  show our measured photometric points from the ACS (filled) and
  H-alpha (open) images. The triangles show photometric points
  obtained from ground based images by Trager et al. The dashed line
  is Trager's Chebychev fit. The solid line is our smooth fit to the
  combination of the ACS points inside 40\arcsec\ and Trager's
  Chebychev fit outside.}
  \label{sb}
\end{figure}

\section{Surface Brightness Profile}

\vspace{10pt}

The surface density profile at large radii for $\omega$ Cen has been
measured from a combination of star counts and aperture photometry
from ground based images \citep{mey87,tra95,van00}. We measure the
central part of the profile taking advantage of
$\it{Hubble~Space~Telescope}$ ($\it{HST}$) spatial resolution. We
measure integrated light from an ACS F435W image (340 sec) applying
the technique described in detail in \citet{noy06}, which uses a
robust statistical estimator, the bi-weight, to calculate number of
counts per pixel on a given annulus around the center of the
cluster. As a test, we also measure the profile from a narrow band
H-alpha image from the Rutgers Fabry-Perot \citep{xie06} with lower
spatial resolution. Since both images have a limited radial coverage,
we use the Chebychev fit of \citet{tra95} for the surface brightness
profile to cover the full, radial extent of the cluster. All profiles
are normalized to the Trager profile.

Having accurate coordinates for the center of the cluster is crucial
when measuring density profiles. Using the wrong center typically
produces a shallower inner profile. We use a technique where we take
an initial guess center, divide the cluster in eight concentric
sectors around this center, and calculate the standard deviation of
the sum of stars for the eight sectors. The radius of the sectors is
chosen to be as large as the image will allow, in this case it is
$\sim$2\arcmin. We repeat the procedure for a grid of center
coordinates and use the one that has the minimum standard
deviation. Details about the technique can be found in
\citet{noy06}. The coordinates for our center are RA $13:26:46.043$ and
DEC $-47:28:44.8$ on the ACS dataset J6LP05WEQ using its WCS zeropoint.

The measured profiles from the B-band and H-alpha images are
consistent as can be seen on Figure \ref{sb}. The H-alpha profile
follows the turnover around the core radius very well up to
100\arcsec~and it also shows the rise toward the center, but it is
noisier than the ACS profile. The solid line is a smooth fit made to
the combination of the photometric points from ACS inside
40\arcsec~and Trager et al. Chebychev fit outside 40\arcsec. For
comparison, we include the \citet{tra95} photometric points in the
plot. The surface brightness profile shows a continuous rise toward
the center with a logarithmic slope of $-0.08\pm0.03$, which is in
contrast to the common notion that $\omega$ Cen has a flat
core. \nocite{van00} Van Leeuwen et al (2000) perform star counts for
giant stars and notice that they are more concentrated than previously
thought. Our result is consistent with their finding.~\citet{bau05}
perform N-body models of star clusters with initial King profiles and
containing a central black hole. They predict the formation of a
shallow cusp of $-0.1$ to $-0.3$ logarithmic slope after $1.5-4$
relaxation times.~\citet{tre07} perform similar N-body models and
conclude that clusters containing central IMBHs evolve to
configurations that have extended cores, with $r_c/r_h\sim0.3$. The
$r_c/r_h$ ratio for $\omega$ Centauri is 0.33, which is in very good
agreement with the predictions from this models. Our observed surface
brightness profile is intriguing considering that it follows the two
specific predictions from these N-body models, but of course, only
kinematical measurements can determine the mass profile, and test the
existence of a central black hole.

\section{Kinematic Measurements}

\vspace{10pt}

Obtaining kinematic information of the central regions of Galactic
globular clusters is a challenging task since the brightest stars
dominate with typical ground-based conditions and the extreme crowding
produces confusion. Measuring individual radial velocities requires a
spatial resolution that can only be achieved with adaptive optics or
from space. On the other hand, measuring velocity dispersion from an
integrated spectrum is subject to shot noise due to giant stars whose
contribution dominates the light. \citet{dub97} calculate the relative
contribution to integrated light by different stellar groups. They
find that the contribution from the few brightest stars is equal in
weight to that of the much larger numbers of fainter stars. Therefore,
the only way to obtain accurate radial velocity dispersion
measurements from an integrated spectrum is if the participation of
the brightest stars can somehow be avoided or at least minimized. One
way to do this is by observing crowded regions with an integral field
unit (IFU) which produces individual spectra of subsections in the
region (typically $\sim$0.2\arcsec\ in size). One can exclude the
spectra affected by the brightest stars when measuring the integrated
background light and thus decrease the shot-noise contribution to the
uncertainty. The Gemini telescopes operate primarily using a queue
scheduling, which makes them an excellent tool to measure integral
field spectra of globular clusters since observing constraints (such
as excellent seeing) can be specified in advance, and data is only
taken when the required observing constraints are met.

\begin{figure}
  \plotone{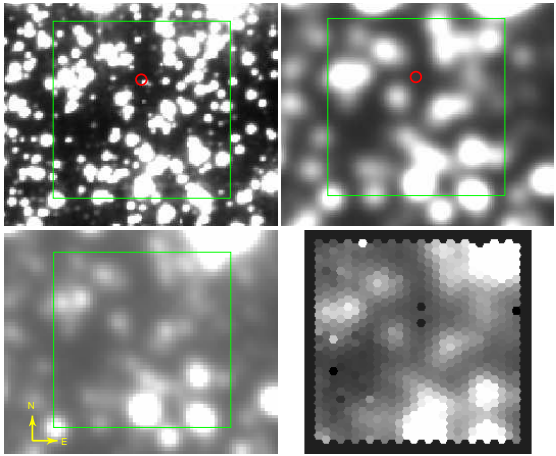}
  \caption{The central field observed with the GMOS-IFU. The green box
  represents the GMOS field of view. Top left: ACS image of the
  observed region. The red circle marks the center of the cluster. Top
  right: Convolved ACS image to reproduce the reported seeing during
  observations. Bottom left: GMOS acquisition image. Bottom right:
  Reconstructed GMOS-IFU image.}
  \label{frame59}
\end{figure}

\begin{figure}
  \plotone{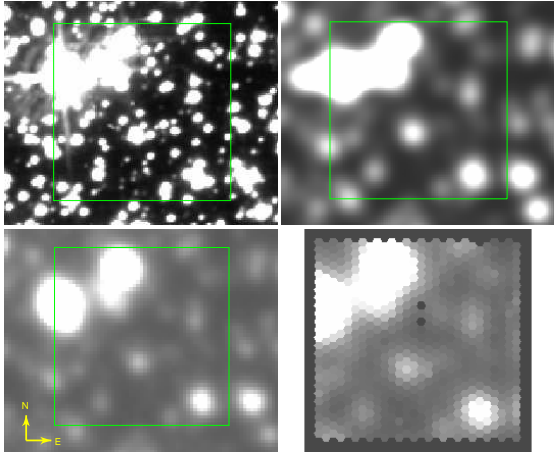}
  \caption{Same as fig 2 for the field 14\arcsec\ away.}
  \label{frame52}
\end{figure}

As part of the Science Verification program for the Gemini GMOS-South
IFU, we obtained nod-and-shuffle observations on February 29 2004
(program ID: GS-2003B-SV-208). We use the IFU in 2-slit
nod-and-shuffle mode, which gives a field of view of
$5\arcsec\times5\arcsec$, comprised of 700 individual lenslets plus
fiber elements, each of which covers approximately 0\arcsec.2 on the
sky. We use the R600 grating, yielding a resolving power R=5560, which
we measure from the lamp spectral lines, along with the Calcium
Triplet filter to give a wavelength coverage of 7900-9300\AA. Three
fields are observed, each for a total integration time of 900 sec on
source and 900 sec on sky. The observations are made using the
nod-and-shuffle technique with 30 sec sub-integrations observed in a
B-A-A-B pattern, where A is the on-source position and B is the sky
position, located 498\arcsec\ away. The nod-and-shuffle technique
improves the sky subtraction, especially in the presence of CCD
fringing, by sampling the object and sky on exactly the same CCD
pixels, with exactly the same light path, on timescales comparable to
those of the sky emission line variability. The first of the three
fields is located at the cluster center, and the second field is
centered 14\arcsec\ away. The third field appears to have been
pointing somewhere else but, despite much efforts, we cannot determine
exactly where the IFU observations are pointed (they do not match
anything in the acquisition image for this field). The reconstructed
IFU image for this third field contains fewer stars and the PSF is
obviously broader than for the other two. It is clear that the
exposure was taken during much worse seeing conditions than the other
two fields, so shot-noise effects are likely to be important; for this
reasons we exclude the third field from further analysis. Using the
standard tasks from the IRAF-GEMINI package we sky subtract,
flat-field, and extract the spectra for each fiber and apply a
wavelength calibration.

\begin{figure}
  \plotone{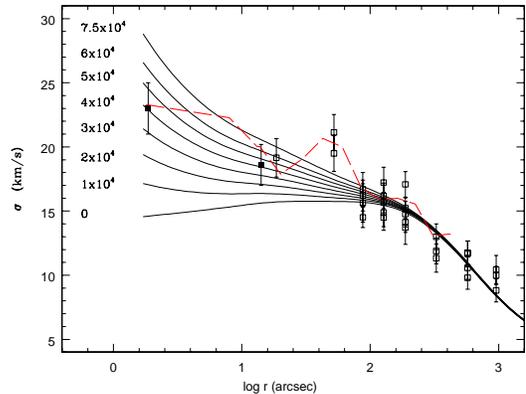}
  \caption{Velocity dispersion profile for $\omega$ Cen with various
  central black hole models. Filled squares are the dispersions and
  uncertainties from the GMOS-IFU and open circles are from individual
  radial velocity measurements. A set of isotropic spherical models of
  varying black hole masses is shown for comparison. The thick line is
  the no black hole model and the thin lines represent models with
  black holes as labeled. The dashed (red) line is the velocity
  dispersion profile for the best-fit orbit-based model (in Section
  4).}
  \label{models}
\end{figure}

The standard flat subtraction does not remove all of the fringing
pattern in the image. As a result, a constant number of counts have to
be subtracted from the data frames before flattening in order to
reduce fringing problems. Relative to the bias frame, the amount of
additional counts from scattered light is about 8\%. Even after this
procedure, there is some residual fringing that can only be alleviated
by combining individual fibers over the full field into one
spectrum. To combine individual fibers we first divide by the
continuum. The intention is to de-weight the bright stars with respect
to the fainter ones, which helps to lessen the problems due to shot
noise. For the continuum fit, we run a boxcar of dimension
$111\times1$ over the reduced image, and then divide the central pixel
by the median of the pixels in the box. This procedure brings all
spectra to the same continuum level. We then combine every individual
fiber with the six adjacent ones, since this represents one seeing
disk for the observations

Figure \ref{frame59} shows the reconstructed image from the IFU fibers
for the central frame and the acquisition image as well as the same
region on the ACS image. We also show a convolved image (with the
reported seeing for the observations) of the ACS frame. The same match
is performed for the field 14\arcsec~away (Fig \ref{frame52}). Both
ACS fields contain $\sim100$ resolved stars. We construct a luminosity
function for the detected stars for each field and compare it to the
luminosity function of the entire cluster core. The luminosity
function is consistent between the two fields. The brightest stars
detected in both fields are two magnitudes fainter than the brightest
stars detected in the core of the cluster. This excludes the
possibility of the integrated spectra being artificially broadened by
the presence of more blue straggler stars in the central field
compared to the field 14\arcsec\ away. Using the photometric
measurements of individual stars together with the reported seeing, we
calculate how many stars contribute to each fiber. Excluding the
fibers which are dominated by a single star we estimate that the
integrated spectrum of the background unresolved light represent about
60 stars in both fields.

We focus on the Ca triplet region (8450\AA -8700\AA) for our
analysis. We measure the relative velocities between each fiber for
the two fields and obtain velocity distributions from the individual
fiber velocities. We fit a Gaussian to the velocity distributions and
observe that the one for the central field is clearly broader than for
the one 14\arcsec\ away. The largest relative velocity between two
fibers is 80 \kms\ for the central frame, and 60 \kms\ for the other
one. Using the dispersion of the individual fiber velocities as a
measure of the cluster velocity dispersion will be biased. Since
multiple stars, in general, provide light to an single fiber, the
measured velocity in that fiber will be pulled toward the cluster mean
as opposed to representing one star. Thus, the dispersions of the
fiber velocities will be biased significantly low. This is what we
find although the central frame does have a obviously larger spread in
fiber velocities.

To properly estimate the velocity dispersion we have to rely on the
integrated light, and require template stars in this
case. Unfortunately, we do not have isolated stars that are free from
the fringe problems mentioned earlier, so we cannot accurately use
template stars observed with the same instrument. We rely on the
template stars observed by \citet{wal05}, from VLT-UVES observations
at around R=35000. We convolve the spectra to our measured resolving
power. To extract the velocity dispersion from the integrated light we
utilize the non-parametric, pixel-based technique as described in
\citep{geb00b,pin03}. We choose an initial velocity profile in bins,
convolve it with the template (or set of templates), and calculate
residuals to the integrated spectrum. The parameters for the line of
sight velocity distribution (either velocity bin values or, if
desired, a parametric Gauss-Hermite expansion) are varied to provide
the best match. Monte Carlo simulations determine the uncertainties,
and use the measured noise in the spectrum.

The dispersion fitting routine allows for a mismatch in the equivalent
width between the object and template. In this case, there is a 30\%
difference in the equivalent width of the calcium triplet lines.  We
do not know whether this is caused by the scattered light (unlikely
given the amplitude), the specific templates we used, omega Cen's
particular composition, or a combination of all three. We have run a
variety of tests to determine whether stars of difference equivalent
widths would cause a bias in dispersions, and find no such bias. We
have also measured the dispersions using template stars from the same
IFU data, since there is at least one star that is fairly
isolated. The uncertainties are larger due to the scattered light
problems, but the value of the dispersion remains the same. Thus, we
conclude that template issues are not a significant source of bias in
the dispersion estimate.

\begin{figure}
  \plotone{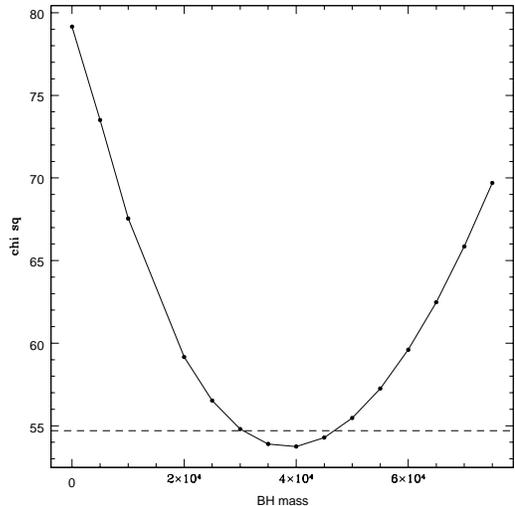}
  \caption{$\chi^2$ vs. black hole mass.  The minimum is found for a
  black hole mass of $4.0\times10^4 M_\odot$, with 68\% confidence
  limit at $3$ and $4.75\times10^4 M_\odot$ marked by the dashed
  line. For our model assumptions, the no black hole model is excluded
  at greater than the 99\% confidence}
  \label{chibh}
\end{figure}
 
We combine the spectra from individual fibers using a biweight
estimator. Different sets of fibers for each frame are combined in
order to test for consistency in our results. First, we combine every
fiber on the frame, then we exclude the 25\%, 50\% and 75\% brightest
fibers. We measure the velocity dispersion for these four spectra for
each frame. The measured velocity for the central frame is always
higher than the one for the frame 14\arcsec\ away for every equivalent
pair of combined spectra. We measure velocity dispersions from 21.8 to
25.2 \kms\ for the central field, and 18.2 to 19.1 \kms\ for the field
14\arcsec\ away. We adopt $23.0\pm2.0$\kms\ for the central field and
$18.6\pm1.6$\kms\ for the other. The latter measurement coincides with
the central velocity dispersion value measured for $\omega$ Cen by
various authors. Van den Ven et al. (2006) measure a line of sight
velocity dispersion profile by combining various datasets
\citep{sun96,may97,rei06,xie06}.  They use 2163 individual radial
velocity measurements divided into polar apertures to obtain the final
velocity dispersion profile. We use dispersion estimates as presented
in their table 4. The average radius for the spectra that contribute
to the central values is 2.5\arcsec\ and 14\arcsec\ for the
second. Figure \ref{models} presents the velocity dispersion data.

\section{Models}

\vspace{10pt}

As discussed in Section 2, the central shape of the surface brightness
profile of $\omega$ Cen resembles that found by \citet{bau05} in star
clusters harboring black holes. The presence of an intermediate black
hole at the center of this cluster is one of the possibilities for
explaining the observed rise in velocity dispersion. We have run two
types of modeling: 1) spherical and isotropic, and 2) flattened,
orbit-based models that allow for general anisotropy. Although the
orbit-based models are more general, they do not consider dynamical
stability. For a system with a half light relaxation time shorter than
its age, like $\omega$ Cen, the dynamical evolution cannot be ignored.
Therefore, while the two models give similar results we quote results
from the isotropic analysis and use the orbit-based models to explore
possible velocity anisotropies.

\begin{figure}
  \plotone{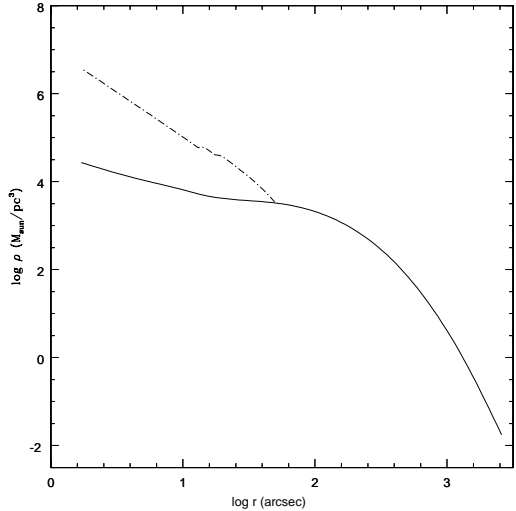}
  \caption{Inferred density profiles. The solid line is the
    deprojected density profile for the luminous component. The dashed
    line represents the required dark component to reproduce the
    observed kinematics.}
  \label{dens}
\end{figure}

For the isotropic analysis, we create a series of models using the
non-parametric method described in \citet{geb95b}. As a first step we
apply a reddening correction to the observed surface brightness
profile. \citet{har96} reports a 0.1 reddening for this cluster,
which, although being relatively low, it is important for the proper
$M/L$ determination of the models. The reddening correction will only
affect the $M/L$ value of the models, but not the shape of the
profiles. We deproject the surface brightness profile using Abel
integrals assuming spheroidal symmetry. The integral involves a
derivative of the profile, therefore, any amount of noise present is
amplified. We apply a spline smoother to the surface brightness
profile before deprojecting and thus obtain a luminosity density
profile as discussed in \citet{geb96}. By assuming an M/L ratio, we
calculate a mass density profile, from which the potential and the
velocity dispersion can be derived. We repeat the calculation adding a
variety of central point masses ranging from 0 to
$7.5\times10^4M_\odot$ while keeping the global $M/L$ value fixed. Van
de Ven et al (2006) measure a fairly constant stellar $M/L$ profile
for $\omega$ Cen of $2.5\pm0.1$. We find a constant mass luminosity
ratio of 2.7, since this provides the best match to the observed
velocity dispersion profile outside the core.

Figure \ref{models} shows the comparison between the different models
and the measured dispersion profile. The most relevant part of the
comparison is the rise inside the core radius, in particular the rise
between the two innermost measurements. As it can be seen, an
isotropic model with no black hole present predicts a slight decline
in velocity dispersion toward the center, instead we observe a clear
rise. The predicted central velocity for the no black hole model is
14.6 \kms\ which is well below any line of sight velocity dispersion
measured inside 1\arcmin. The calculated $\chi^2$ values for each
model are plotted in Figure \ref{chibh}, as well as a line showing a
$\Delta\chi^2=1$. The $\chi^2$ curve implies a best-fitted black hole
mass of $4^{+0.75}_{-1}\times10^4M_\odot$. Even the original velocity
dispersion profile without our two innermost measurement already
points to an intriguing discrepancy, but the central measurements
confirm an important rise in $M/L$ from the core radius to the center
of this cluster. The central $M/L$ value is 6.7, which is a
considerable rise from the value of 2.7 just inside the core
radius. Our best fit model implies a central density of
$5.6\times10^7M_\odot/pc^3$ the largest measured in a globular
cluster.

\begin{figure}
  \plotone{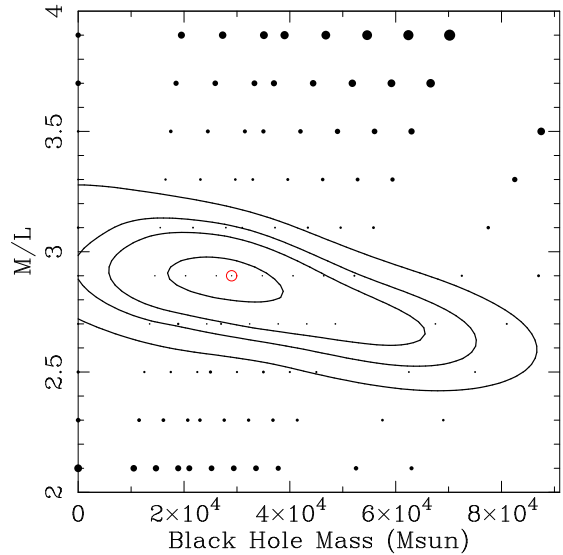}
  \caption{Contours of $\chi^2$ as a function of black hole mass and
mass-to-light ratio. Each point represents a particular model. The
contours represent the 68, 90, 95, and 99\% confidence for one
degree-of-freedom, implying $\Delta\chi^2=1.0, 2.7, 4.0,$ and 6.6. The
circled point is the model that has the minimum value.}
  \label{2dchi}
\end{figure}

We also construct axisymmetric orbit-based dynamical models. The
models are based on the formulation by \citet{sch79} and are
constructed as in \citet{geb00b,geb03}. These models provide the most
freedom possible of the distribution function for an axisymmetric
system. We use the same deprojection as describe above, except we also
include the observed flattening. Assuming an $M/L$ ratio and a BH
mass, the mass distribution of the cluster is obtained and from it,
the potential can be computed. Using this potential, we generate about
$10^4$ representative orbits. The best match to the observed
photometric and kinematical data provide the orbital weights for a
given potential. The process is repeated for various $M/L$ values and
BH masses until the minimum $\chi^2$ model is found. The kinematical
observations matched here are the individual radial velocities from
van den Ven et al. (2006), and the two integrated measurements
described above. We plan to perform more detailed orbit-based models
including proper motion measurements in the future.

Figure \ref{2dchi} plots $\chi^2$ as a function of black hole mass and
stellar M/L. The best fit model requires a black hole mass of
$3\pm1\times10^4~M_\odot$, with 1-sigma of the isotropic result. As
expected, since orbit-based models are more general, the difference in
$\chi^2$ compared to the no black hole case is smaller than in the
isotropic case. The $\Delta\chi^2$ between the best-fitted black hole
model and no black hole model is 4 (marginalized over M/L), implying a
95\% significance. In comparison, the $\chi^2$ difference for the
isotropic models is 25. As discussed below, the reason for the
difference is that the orbit-based models allow for significant radial
anisotropy in the core for the no black hole model, while the best-fitted
mass produces a nearly isotropic distribution consistent with previous
measurements \citep{ven06}.

We also plot the velocity dispersion profile as measured in the
best-fit orbit based model in Figure \ref{models}, given as the dashed
(red) line. The dispersion profile for the orbit-based model with no
black hole is very similar, with no obvious correlated
differences. The reason there are no obvious differences is because
the overall change in $\chi^2$ is small and the orbit-based models
tend to redistribute orbital weights to spread $\chi^2$ over the full
radial range. As discussed in Gebhardt et al. (2003), it is difficult
to see differences in radial profiles of projected kinematics between
models with black holes and without, even for galaxies where the
overall $\chi^2$ difference is large. This is understood since the
orbit models have the freedom to change the orbital properties in such
a way to even out the $\chi^2$ differences over the full spatial
extent of the observations. For omega Cen, the black hole model only
provides an increase in $\chi^2$ of 4; thus with 23 data points, the
average difference in terms of the measurement uncertainty is 0.42. For
these reasons, the orbit-based models provide a modest significance
for a black hole, and the argument is significantly strengthened when
considering the need for the strong amount of radial anisotropy when a
black hole is not included.

The observations we use in the dynamical modeling rely on only the
first and second moments of the velocity distribution. Since we have
individual velocities (except for the central two radial points), we
can utilize the full velocity profile. Furthermore, proper motion data
exists for data at larger radii (van den Ven et al. 2006). Including
both effects, full velocity profiles and proper motions, will be the
subject of a more detailed paper on omega Cen.

\subsection{Alternative to a Black Hole: Dark Remnants}

A possible alternative to explain the observed rise in $M/L$ toward
the center is a concentration of dark stellar remnants, e.g.  neutron
stars, stellar mass black holes, or massive white dwarfs. Using the
observed velocity dispersion profile, we calculate the total enclosed
mass and from this, the mass density profile. We then compare this
with the enclosed mass implied by the luminosity density profile,
assuming the same M/L as for the models in the previous section. From
these two profiles, we can estimate the density profile of the implied
extended dark component, if we assume this was the cause of the
velocity dispersion rise towards the center. Figure \ref{dens} shows
the estimated density profile for the dark and luminous components. It
is clear that, if the velocity rise is due to an extended distribution
of dark stellar remnants, the density profile of this dark component
needs to be very concentrated and steep, with a logarithmic slope of
$\sim-2.0$ (the slope of the dashed line in the figure), resembling a
cluster undergoing core collapse. The relaxation time for $\omega$ Cen
implies a much slower dynamical evolution than the one necessary to
reach such a configuration.  Core-collapse models have shown that when
a cluster has reached such a high degree of mass segregation, the
observable core to half light radius gets very small, with values
below 0.05 \citep{bre94,mak96}, while this ratio is 0.3 for
$\omega$~Cen. Also, the concentration value for $\omega$~Cen is
$c=1.6$ \citep{har96}, which is too low a value for the cluster to
have undergone core-collapse. No evolutionary model predicts such a
concentrated distribution of dark remnants inside a cluster with a
shallow extended core for the visible stars.  Furthermore, the
required number of dark remnants is around 1\% of the cluster mass.
While this number is expected from stellar evolution, it is not
expected to have all of the remnants to be concentrated inside of the
core.

\begin{figure}
  \plotone{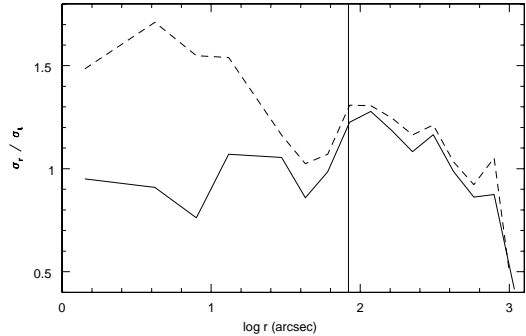}
  \caption{Radial over tangential anisotropy vs. radius from
    orbit-based models. The solid line is for the best fit model
    containing a black hole. The dashed line is for a model with no
    black hole present. The vertical line marks the location of the
    core radius.}
  \label{anis}
\end{figure}

\subsection{Alternative to a Black Hole: Orbital Anisotropy}

Another possibility is that the observed rise in velocity dispersion
is due to velocity anisotropy in the cluster. The orbit-based models
explore this possibility. The two main results from the orbit analysis
are that 1) the lowest $\chi^2$ model is the one with a central black
hole and 2) the model with no black hole requires a substantial amount
of radial anisotropy. Figure \ref{anis} shows that without the
presence of a central black hole, a large degree of radial anisotropy
-- $\sigma_r/\sigma_t$=1.5 -- is required inside $0.3~r_c$. At
$r>28$\arcsec, the models with and without a black hole are close to
isotropic, in agreement with the results of van den Ven et
al. (2006). For a system as dense as $\omega$~Cen, such a degree of
anisotropy as measured in the no black hole case is expected to be
quickly erased through relaxation processes. However, even with such a
strong amount of radial anisotropy, the no black hole case is a poorer
fit than the best-fitted black hole. One of the disadvantages of the
orbit-based models is that we cannot include the dynamical stability
arguments in the $\chi^2$ analysis, and for a system with a short
relaxation time such as a globular cluster, this may be important. We,
therefore, adopt results from the isotropic models, in particular
since the analysis of van den Ven et al. (2006) find an isotropic
distribution.

\section{Discussion}

\vspace{10pt}

We measure the surface brightness profile for the globular cluster
$\omega$ Centauri (NGC~5139) from an ACS image in the central
40\arcsec. The profile shows a continuous rise toward the center with
a logarithmic slope of $-0.08\pm0.03$, in contrast with previous
measurements which found a flat core. The shape of the profile is
similar to that obtained from numerical models of star clusters
containing black holes in their centers. We measure a line of sight
velocity dispersion for two 5\arcsec$\times$5\arcsec~regions, one at
the center of the cluster and the other 14\arcsec~away. We detect a
rise in velocity dispersion from 18.6 \kms\ for the outer field to 23
\kms\ for the central one. We combine these two measurements with
previously measured velocity dispersion at larger radii.

When we compare the observed velocity dispersion profile with a series
of isotropic models containing black holes of various masses, we find
that a black hole of $4.0 ^{+0.75}_{-1.0}\times~10^4 M_\odot$ is
necessary to match the observations. We explore alternative
explanations for the observed rise in our central velocity dispersion
measurements. First we consider the possibility that the observed
$M/L$ rise is due to the presence of an extended component composed of
dark remnants such as neutron stars or faint white dwarfs. The density
profile of the dark component is required to be extremely concentrated
toward the center, with a configuration practically decoupled from the
luminous component. $\omega$ Cen has a weak cusp in the central
luminosity density profile, implying that the gravitational potential
is very shallow inside the core and therefore mass segregation is only
a weak effect. \citet{fer06} confirm the lack of segregation by
measuring the radial distribution of blue straggler stars, which are
heavy stars and should sink to the center of the cluster if there is
mass segregation. They find a flat radial distribution of blue
stragglers with respect to lighter stellar populations. Also, the
formation channel of the blue stragglers is not collisional, as it
would be expected if there was a considerable amount of mass
segregation. With this evidence in hand, there is no reason to expect
a large variation of $M/L$ inside the core due to stellar content, so
a detected rise in $M/L$ is likely to come from the presence of a
concentrated massive object.

\begin{figure}
  \plotone{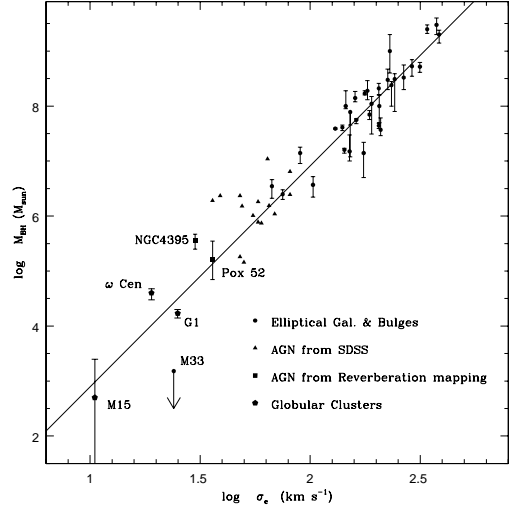}
  \caption{$M_\bullet-\sigma_{vel}$ relation for elliptical galaxies
  and bulges. The solid line is the relation in \citet{tre02}.
  $\omega$ Cen lies on the low mass extrapolation and suggests a
  similarity between it and the galaxies. Different types of systems
  such as star clusters and low luminosity AGN appear to populate the
  low mass end of the diagram.}
  \label{bhsigma}
\end{figure}

There is also a stability argument against a dense compact cluster of
dark remnants. The central density as measured from the 23~\kms\
dispersion estimate at 1.8\arcsec\ is $5.6\times10^7
M_\odot/pc^3$. This is the largest measured for a globular cluster and
it would be difficult to maintain using stellar remnants. Obviously,
if the density is due to solar mass remnants, over $10^4$ remnants
would be required inside of 0.05 parsecs. Using the arguments of Maoz
(1998) and Miller (2006), this mass and density makes $\omega$~Cen one
of the better examples where stellar remnants can be ruled out due to
evaporation. Maoz estimates that for these numbers, any cluster of
remnants will have evaporated within the age of the cluster.

An observed velocity dispersion rise toward the center of a cluster
can also occur if a degree of anisotropy is present. If more radial
orbits are present, those stars pass near the center at higher
velocities than they would in an isotropic case. This possibility is
evaluated with our orbit-based models. Our results agree with the
model by van den Ven et al. (2006) in showing no anisotropy in the
central 10 arcmin. Their models show a degree of tangential anisotropy
at large radius, but no radial anisotropy. Models without a central
black hole but having a large degree of anisotropy inside 28\arcsec\
are not as as good as models including a black hole. Furthermore, the
high degree of radial anisotropy is highly unstable in dense systems
like $\omega$~Cen and therefore it is an unlikely explanation for the
observed kinematics.

Figure \ref{bhsigma} shows the known $M_\bullet-\sigma_v$ relation for
black holes in elliptical galaxies and bulges \citep{geb00a,fer00}. The
galaxies used to determine the relation \citep{tre02} are plotted
along with objects containing smaller black holes in low luminosity
quasars \citep{bar05}, two nearby low luminosity AGN (NGC~4595 and
Pox~52), and three globular clusters (G1, M15 and $\omega$ Cen). We
also plot the upper limit for the black hole mass in the nucleus of
M33 \citep{geb01}, which does not lie on the correlation. The black
hole in $\omega$ Cen lies above the relation, but it is consistent
with the scatter observed a larger masses.  The measured black hole
mass is 1.6\% of the total mass of the cluster, which is much larger
than the canonical value of $\sim0.3$\% for larger spheroids
\citep{mag98}. If $\omega$ Cen is indeed the nucleus of an accreted
galaxy it is expected that it's original mass was considerably larger
than what we measure now. \citet{bek03} reproduce the current mass and
orbital characteristics of $\omega$ Cen with a model of an accreted
$10^7M_\odot$ dwarf galaxy. A mass of $4\times10^7M_\odot$ for the
original spheroid would put the black hole near the 0.3\% value.

The two pieces of observational evidence that $\omega$ Cen could
harbor a central black hole come from the photometry and the
kinematics. From the HST image of $\omega$ Cen, we measure a central
logarithmic surface brightness slope of $-0.08\pm0.03$. This value is
very similar to that claimed by the N-body simulations of Baumgardt et
al. (2005) that are most likely explained by a central black hole.
Standard core-collapse does not lead to such a large core with a
shallow central slope. The black hole tends to prevent core collapse
while leaving an imprint of a shallow cusp. It will be important to
run models tailored to $\omega$~Cen to see if one can cause and
maintain a shallow cusp without invoking a central black hole.
However, the main observational evidence for the central mass comes
from the increase in the central velocity dispersion, where we detect
a rise from 18.6 to 23~\kms\ from radii of 14 to 2.5\arcsec. In fact,
even excluding the Gemini data presented here, the previous
ground-based data suggest a central mass concentration as well. The
core of $\omega$~Cen is around 155 \arcsec\ (about 2.5 \arcmin), so
the dispersion rise is seen well within the core.

\acknowledgments

\vspace{5pt}

E.N. would like to thank Tim de Zeeuw for very valuable discussions
and Glenn van de Ven for kindly sharing his data. K.G. acknowledges
NSF CAREER grant AST 03-49095. We thank Carl Jakob Walcher for
promptly making his data available to us. This publication is based on
observations made with the NASA/ESA Hubble Space Telescope, which is
operated by the Association of Universities for Research in Astronomy,
Inc., under NASA contract NAS 5-26555, and observations obtained at
the Gemini Observatory, which is operated by the Association of
Universities for Research in Astronomy, Inc, under cooperative
agreement with the NSF on behalf of the Gemini partnership: the
National Science Foundation (United States), the Particle Physics and
Astronomy Research Council (United Kingdom), the National Research
Council (Canada), CONACYT (Chile), the Australian Research Council
(Australia), CNPq (Brazil) and CONICET (Argentina). We acknowledge the
technical support from the Canadian Astronomy Data Centre, which is
operated by the Herzberg Institute of Astrophysics, National Research
Council of Canada, and the support by CONACYT.

\end{document}